\documentclass[usenatbib]{mn2e}

\usepackage{epsfig}
\usepackage{mathrsfs,aas_macros}

\def\Lya{Ly$\alpha$~}

\def\HI{\hbox{H$\,\rm \scriptstyle I\ $}}
\def\HII{\hbox{H$\,\rm \scriptstyle II\ $}} 
\def\HeI{\hbox{He$\,\rm \scriptstyle I\ $}}
\def\HeII{\hbox{He$\,\rm \scriptstyle II\ $}}
\def\HeIII{\hbox{He$\,\rm \scriptstyle III\ $}}

\title[Thermal constraints on hydrogen reionisation]{Thermal constraints on the
  reionisation of hydrogen by population-II stellar sources}

\author[S. Raskutti et al.] {Sudhir Raskutti$^{1}$, James
  S. Bolton$^{1}$, J. Stuart B. Wyithe$^{1,3}$ \& George D. Becker$^{2}$ \\ $^1$ School of Physics, University of
  Melbourne, Parkville, VIC 3010, Australia \\ $^2$ Kavli Institute
  for Cosmology and Institute of Astronomy, Madingley Road, Cambridge,
  CB3 0HA \\ $^3$ ARC Centre of Excellence for All-sky Astrophysics 
  (CAASTRO)}

\begin{document}

\date{}

\maketitle

\label{firstpage}

\begin{abstract}
Measurements of the intergalactic medium (IGM) temperature provide a
potentially powerful constraint on the reionisation history due to the
thermal imprint left by the photo-ionisation of neutral hydrogen.
However, until recently IGM temperature measurements were limited to
redshifts $2 \leq z \leq 4.8$, restricting the ability of these data
to probe the reionisation history at $z>6$.  In this work, we use
recent measurements of the IGM temperature in the near-zones of seven
quasars at $z\sim 5.8 - 6.4$, combined with a semi-numerical model for
inhomogeneous reionisation, to establish new constraints on the
redshift at which hydrogen reionisation completed.  We calibrate the
model to reproduce observational constraints on the electron
scattering optical depth and the \HI photo-ionisation rate, and
compute the resulting spatially inhomogeneous temperature distribution
at $z\sim6$ for a variety of reionisation scenarios.  Under standard
assumptions for the ionising spectra of population-II sources, the
near-zone temperature measurements constrain the redshift by which
hydrogen reionisation was complete to be $z_{\rm r} > 7.9$ $(6.5)$ at
68 (95) per cent confidence.  We conclude that future temperature
measurements around other high redshift quasars will significantly
increase the power of this technique, enabling these results to be 
tightened and generalised.

\end{abstract}
 


\section{Introduction}

The reionisation of cosmic hydrogen represents a significant moment in
the history of the Universe. The appearance of the first stars and
quasars heated and ionised the IGM, catalysing
the transition in which intergalactic hydrogen changed from being
predominantly neutral to its present day, highly ionised
state. Understanding exactly how and when this phase transition
occurred will offer insight into the nature of the first
sources of light (e.g. \citealt{Barkana2001}). However, existing
constraints on the timescale and extent of reionisation are currently
uncertain (\citealt{ChoudhuryFerrara2006, Pritchard2009, Mitra2011}).

There are two primary observations which are used to constrain the
epoch of hydrogen reionisation.  The first of these are the absorption
spectra of quasars at $z\simeq 6$
(\citealt{White2003,Fan2006,Becker2007}).  These enable detailed study
of the IGM ionisation state at high redshift, and indicate that the
Universe was reionised at $z \geq 6$ (\citealt{GnedinFan2006}, but see
also \citealt{Mesinger2010}). The second observation is provided by
measurements of the Thomson optical depth, ${\tau}_{\rm CMB}$, to the
surface of last scattering, which indicate that hydrogen reionisation
could not have ended any earlier than $z_{\rm r} = 10.6 \pm 1.2$
(\citealt{Komatsu2011}).  In practice, this constraint is consistent
with a wide variety of extended reionisation scenarios.

Analytical and semi-numerical models have been confronted with a
combination of these and other observations to place more quantitative
limits on the reionisation redshift. In particular, studies by
\citet{ChoudhuryFerrara2006}, \citet{Pritchard2009} and
\citet{Mitra2011} employ the above observables, alongside the observed
distribution of Lyman limit systems and constraints on the star
formation rate, to model reionisation. These models can be used to
infer constraints on the extent and completion time of reionisation.
However, the lack of direct observational evidence at $z > 6$ means
that parameters in these models, such as the ionising background
intensity and the thermal state of the IGM, are relatively uncertain
at high redshift. This implies that, from current observations alone,
constraints on the redshift by which reionisation is complete remain
loose, and cannot be improved beyond $6 < z_{\rm r} < 10.5$
(\citealt{Mitra2011}).

A promising additional approach comes from measurements of the IGM
temperature (\citealt{Zaldarriaga2001, Trac2008, FurlanettoOh2009,
  Bolton2010}).  The relatively long time scale of adiabatic cooling
implies that the low redshift IGM still carries the thermal imprint of
reionisation heating from much earlier times. Several authors have
previously used this fact, combined with temperature measurements at
$z < 4$, to constrain the end of the hydrogen reionisation epoch to
$z_{\rm r} < 10$ (\citealt{Theuns2002a,
  HuiHaiman2003}). Unfortunately, because the imprint of reionisation
heating at $z_{\rm r} > 10$ reaches a thermal asymptote by $z \approx
4$ these studies are limited by the relatively low redshift at which
the temperature measurements were made.

Measurements of IGM temperature at higher redshift therefore offer a
new opportunity to tighten these constraints.  A first step towards
this was made by \cite{Bolton2010}, who obtained a measurement of the
IGM temperature at mean density from \Lya absorption lines in the
near-zone of the $z=6.02$ quasar SDSS J0818$+$1722.  Although the
statistical uncertainty on their single line-of-sight measurement was
large, \cite{Bolton2010} were nevertheless able to use a simple
cooling argument to infer a reionisation redshift of $z_{\rm r} < 9$
(to 68 percent confidence) within the quasar near-zone, assuming
hydrogen was reionised by population-II stars.  Recently,
\cite{Bolton2011a} have significantly extended these observational
constraints by measuring the IGM temperature in the near-zones of six
further quasars at $5.8 \leq z \leq 6.4$.  In this work, we therefore
seek to extend the modelling of \cite{Bolton2010} and include all seven
near-zone temperature measurements to obtain a limit on the {\it
  reionisation redshift for the general IGM.}  Specifically, by using
a semi-numerical model to follow inhomogeneous reionisation and
photo-heating of intergalactic hydrogen by stellar sources, as well
any subsequent photo-heating by the quasar itself, we may calculate
near-zone temperatures as a function of spatial position and
reionisation redshift.  A comparison of our model with the
observational data then allows us to constrain the redshift at which
reionisation completed.

The paper is organised as follows. We begin in
Section~\ref{Sec:Semi-Numerical Model} by introducing our
semi-numerical model of reionisation, and proceed in
Section~\ref{Sec:Nion Models} to give a more detailed description of
the modelling of ionising sources. Predicted temperatures from the
model for different reionisation scenarios are then described in
Section~\ref{Sec:Results}, and we compare these to the
\cite{Bolton2011a} quasar temperature measurements. Our constraints on
reionisation history are then discussed in
Section~\ref{Sec:Constraints}, where we also assess the prospects for
improving on these constraints in the future using additional quasar
sight-lines. Conclusions and further directions are then presented in
Section~\ref{Sec:Conclusions}. Throughout the paper we use
{$\Lambda$CDM} cosmological parameters $\Omega_{\rm m}=0.27$,
$\Omega_{\Lambda}=0.73$, $\Omega_{\rm b}h^{2}=0.0225$, $h=0.704$,
$\sigma_{8}=0.81$, $n_{\rm s}=0.96$, consistent with recent studies of
the cosmic microwave background (\citealt{Komatsu2011}).


\section{Semi-Numerical Reionisation Model}
\label{Sec:Semi-Numerical Model}

Our model of inhomogeneous hydrogen reionisation simulates both the
evolving ionisation state of the IGM and its thermal history.  The
former is calculated using the semi-numerical model developed by
\cite{WyitheLoeb2007} and \cite{geil2008} (see also
\citealt{Zahn2007,Mesinger2007,Choudhury2009,Thomas2009} for similar
approaches). The ionisation state derived from this model is then used
to evaluate both the ionising flux and the mean free path at the Lyman
Limit, and to infer a temperature distribution for the IGM.

\subsection{Ionisation model}
\label{SubSec:Ionisation Model}

We simulate the IGM inside a periodic, cubic grid of side-length $L =
100\,{h^{-1}\,\rm Mpc}$, containing $N = 256$ voxels per side-length,
with an over-density field ${\delta}(x)$ generated using the transfer
function of \cite{EisensteinHu1999}. The calculations detailing the
evolution of ionisation structure in this density field (which is
initialised to a neutral state at $z = 99$) are described in
\cite{geil2008}. Briefly, the model evaluates the ionisation fraction
$Q_{\delta_R,R}$ of a spherical region of scale $R$ around a given
voxel at redshift $z$ as

\begin{eqnarray}
\label{Eq:History}
\nonumber
\frac{dQ_{\delta_R,R}}{dt} &=& \frac{N_{\rm ion}}{0.76}\left[Q_{\delta_R,R} \frac{dF_{\rm col}(\delta_R,R,z,M_{\rm ion})}{dt} \right.\\
\nonumber
&+& \left.\left(1-Q_{\delta_R,R}\right)\frac{dF_{\rm col}(\delta_R,R,z,M_{\rm min})}{dt}\right]\\
&-&\alpha_{\rm B}Cn_{\rm H}^0\left[1+\delta_R\frac{D_1(z)}{D_1(z_{\rm obs})}\right] \left(1+z\right)^3Q_{\delta_R,R},
\end{eqnarray}

\noindent
where ${\delta_R}$ is the region's over-density, $\alpha_{\rm B}$ is
the case-B recombination coefficient and $C$ is the clumping factor.
In this work we shall assume $C=3$, broadly consistent with recent
results from hydrodynamical simulations (\citealt{Pawlik2009,
  RaicevicTheuns2010}). The justification for this choice, as will be
discussed in more detail in Section \ref{SubSec:Current}, is that it
tends to provide the weakest overall constraints on reionisation
redshift.  Meanwhile, $N_{\rm ion}$, the number of ionising photons
entering the IGM per baryon in galaxies, is left as a free parameter,
the modelling of which will be described in more detail in Section
\ref{Sec:Nion Models}.

The production rate of ionising photons is then assumed to be
proportional to $N_{\rm ion}$ and the rate of change of the collapsed
fraction, $F_{\rm col}$, in haloes above some minimum threshold mass
for star formation ($M_{\rm min}$ in neutral and $M_{\rm ion}$ in
ionised regions, corresponding to virial temperatures of $10^4$\,K and
$10^5$\,K respectively, e.g.  \citealt{Dijkstra2004a}). The collapsed
fraction in a region of comoving radius $R$ and mean overdensity
$\delta(z)=\delta(z_{\rm obs}) D_1(z)/D_1(z_{\rm obs})$ is found using the extended
Press-Schechter model \citep{Bond1991}, which gives

\begin{eqnarray}
F_{\rm col}(\delta_R,R,z) = \mbox{erfc}{\left[\frac{\delta_c-\delta_R(z)}{\sqrt{2(\sigma^2_{\rm gal}-\sigma_R^2)}}\right]},
\end{eqnarray}

\noindent
where $\mbox{erfc}(x)$ is the complimentary error function,
$\sigma_R^2$ is the variance of the overdensity field smoothed on a
scale $R$ and $\sigma^2_{\rm gal}$ is the variance of the overdensity
field smoothed on a scale $R_{\rm gal}$, corresponding to a mass scale
of $M_{\rm min}$ or $M_{\rm ion}$ (both evaluated at redshift $z$).

Equation~(\ref{Eq:History}) is solved at every voxel on a range of
scales from $L$ to $L/N$ at logarithmic intervals of width $\Delta R/R
= 0.1$. A voxel is then deemed to be ionised if it lies inside a
region for which $Q_{\delta_R,R}>1$ on any scale $R$. By repeating
this procedure at successive redshifts, we obtain our ionisation field
as a function of $z$.

\subsection{The ionising emissivity}
\label{Luminosity Density}

In order to compute the heating of the IGM, we must next obtain the
ionising emissivity in each voxel of our simulation. We simulate the
ionisation field at 30 regular redshift intervals between an initial
redshift, $z_{\rm 0}$, and the reionisation redshift, $z_{\rm
  r}$. These are, respectively, the redshifts corresponding to $Q =
0.01$ and $Q = 1.0$, where $Q = Q_{\delta_R,R}$ for the case where
equation~(\ref{Eq:History}) is solved at mean density and $R
\rightarrow \infty$.  At each interval, if a voxel is flagged as
ionised, we compute the intrinsic emissivity as the product of $N_{\rm
  ion}$ and the collapsed fraction above a minimum halo mass, $M_{\rm
  ion}$. Using this, the proper ionising photon production rate
($\dot{N_{\rm {\gamma}}}$) in units of photons ${\rm
  cm^{-3}}{\,\rm s^{-1}}$, at each voxel, is

\begin{equation}
\label{Eq:Ndot1}
\dot{N_{\rm {\gamma}}} = \frac{{\rho}_{\rm c}{\Omega}_{\rm b}(1 + z)^{3}(1 + \delta_{\lambda})N_{\rm ion}}{{\mu}m_{\rm p}}\frac{dF_{\rm col}(\delta_{\lambda},{\lambda_{\rm HI}},z,M_{\rm ion})}{dt},
\end{equation}

\noindent
where ${\rho}_{\rm c}$ is the present day critical density, $m_{\rm
  p}$ is a proton mass and ${\mu}$ is the mean molecular weight. In
this instance, the smoothing scale $\lambda$ used to calculate the
collapsed fraction is given by the minimum of the ionised bubble
radius and the mean free path at the Lyman limit, ${\lambda_{\rm
    HI}}$. This is related to the ionising emissivity
$\epsilon_{\nu}$, (in units of $\rm erg\,s^{-1}\,cm^{-3}\,Hz^{-1}$)
above the Lyman Limit frequency, ${\nu_{\rm HI}}$, by

\begin{equation} 
\label{Eq:Ndot2}
\dot{N_{\rm {\gamma}}} = \int_{\nu_{\rm HI}}^{\infty}
\frac{\epsilon_{\nu}}{h_{\rm p}\nu}\, d\nu,
\end{equation}
where we adopt a simple power law spectrum

 \begin{equation} 
\epsilon_{\nu} = \left(\frac{\nu}{\nu_{\rm HI}}\right)^{-\alpha} \times \cases{\epsilon_{\rm HI} & ($\nu_{\rm HI} \leq \nu <
  \nu_{\rm HeII}$),\cr \noalign{\vskip3pt} 0 & ($\nu_{\rm
    HeII} \leq \nu$).\cr}
\label{Lspectrum}
\end{equation}
Here $\epsilon_{\rm HI}$ is the emissivity at the hydrogen Lyman
limit. Below the \HeII ionisation edge, we take a spectral index of
${\alpha} = 3$ characteristic of soft, population-II stellar sources
({\it e.g.}  \citealt{HuiHaiman2003,FurlanettoOh2009}). Above
${\nu}_{\rm HeII}$, the spectrum is zero in order to enforce the
requirement that there be no significant stellar ionisation of He~$\rm
\scriptstyle II$. This is necessary for consistency with our
semi-numerical reionisation model which only traces the propagation of
hydrogen ionisation fronts, and so cannot predict the larger, extended
\HeIII fronts resulting from harder \HeII ionising photons. On the
other hand this is not an unrealistic requirement, since the double
reionisation of helium is thought to not complete until $z\sim3$ when
the number density of quasars is at its peak
(\citealt{MadauMeiksin1994,WyitheLoeb2003,FurlanettoOh2008}). Nevertheless,
this assumption rules out the consideration of more exotic ionisation
models involving early populations of quasars or population-III
sources, which could produce different thermal histories.  More
detailed modelling, which accounts for doubly ionised helium may
therefore be performed in the future.  In this work, however, we
consider {\it hydrogen reionisation by population-II stellar sources
  only}.

\subsection{The ionising background specific intensity}
\label{UV Background Flux}

Having computed the intrinsic emissivity, the specific intensity,
$J_{\rm {\nu}}$, at each voxel may then be calculated by accounting
for the propagation and attenuation of radiation.  This is a
complicated calculation since, in general, it depends on the topology
of \HII regions around each point in the simulation and the
distribution of Lyman limit systems.  A full numerical evaluation
therefore requires radiative transfer, which is not included in our
model. Fortunately, the long-term thermal evolution (for timescales
longer than the local photo-ionisation timescale) is relatively
insensitive to the exact amplitude of $J_{\rm {\nu}}$, and instead
depends primarily on the spectral shape of the ionising radiation
(\citealt{HuiHaiman2003,Bolton2009}).  Therefore, to estimate the
specific intensity we make the simplifying assumption that the mean
free path is much smaller than the horizon scale, and use the local
source approximation (e.g. \citealt{Faucher2009}).  We compute the mean
free path in two regimes, prior to and following the overlap of \HII
regions.

Prior to overlap, the mean free path is set by the size of \HII
bubbles (\citealt{GnedinFan2006}), and may therefore be found by
partitioning the ionisation field into approximately spherical \HII
regions. To do this, we smooth over the ionisation field at successive
scales, employing the same logarithmic descent as for our density
field. At each scale we calculate the size of discontiguous regions
with a standard friend-of-friends (FoF) algorithm. The bubble size for
each voxel is then the scale of the smallest FoF region of which it is
part, augmented by the smoothing scale. This has the effect of
dividing bubbles that only barely overlap into separate spheres so
that the spherical approximation remains valid to a filling factor of
around $Q \approx 0.5$.  We then assume that the optical depth to
ionising photons is infinite beyond the \HII region radius.  Lyman
Limit photons therefore travel one mean free path, corresponding to
the \HII region size, before being absorbed.

In the post-overlap case, the IGM is fully ionised and the mean free
path of \HI ionising photons is constrained by the abundance of Lyman
limit systems rather than the sizes of \HII regions. In this instance,
we employ a fit to the mean-free-path at the Lyman limit in proper
Mpc, taken from \cite{Songaila2010}:

\begin{equation}
\label{Eq:Mfp}
{\lambda_{\rm HI}} = \frac{88.6}{\Gamma(2 - \beta)}{\left(\frac{1 +
      z}{4.5}\right)}^{-4.44} {\rm Mpc},
\end{equation}

\noindent
where $\Gamma$ denotes the gamma function, and ${\beta}$ is the power
law exponent of the \HI column density distribution. {Observationally,
the value of this exponent is unconstrained at $z > 6$, however at
lower redshift measurements indicate $\beta \approx 1.5$
\citep{Petitjean1993, Hu1995}.  Since this is roughly constant over
the range $2 < z < 4$ \citep{Misawa2007}, we simply extrapolate to
high redshift and adopt $\beta = 1.5$ throughout. We note that the
column density distribution will change significantly as we enter the
reionisation epoch, however as we are primarily interested in the
value of $\beta$ in the post-overlap regime, this does not affect our
results.} Finally, we assume that a given \HII region is in the
pre-overlap regime if its radius is less than ${\lambda_{\rm HI}}$
(i.e. the mean free path is not set by Lyman limit systems) and is in
the post-overlap regime otherwise.

Adopting the local source approximation, the ionising background
specific intensity in each voxel may then be approximated by

\begin{equation} J_{\nu} \simeq \frac{1}{4\pi}\epsilon_{\nu}\lambda_{\nu}, \label{Eq:Flux} \end{equation}

\noindent
where $\lambda_{\nu}=\lambda_{\rm HI}(\nu/\nu_{\rm HI})^{3(\beta-1)}$. 
Prior to overlap $\lambda_{\rm HI}$ is fixed to the \HII region
size, while post overlap it is given by equation~(\ref{Eq:Mfp}). Lastly, note that 
in both regimes at frequencies above the Lyman Limit, we have $J_{\nu} \propto
\nu^{-\alpha+ 3(\beta - 1)}$. This effectively hardens the intrinsic
stellar spectrum by $\alpha \rightarrow \alpha - 1.5$ , accounting for
the additional heating due to filtering of the ionising radiation by
discrete, Poisson distributed absorbers in the intervening IGM
(e.g. \citealt{Miralda2003,Faucher2009}).

\subsection{The thermal evolution}
\label{Thermal Evolution}

Once the specific intensity at a given voxel is known, the
photo-ionisation rates, $\Gamma_{\rm i} \rm\, [s^{-1}]$, and
photo-heating rates, $g_{\rm i}\rm\, [erg\,s^{-1}]$, for $i$=H~$\rm
\scriptstyle I$, He~$\rm \scriptstyle I$ and He~$\rm \scriptstyle II$,
may be calculated via:

\begin{equation}
\label{Eq:Gamma}
{\Gamma_{\rm i}} = \int_{\nu_{\rm i}}^{\infty}\frac{4{\pi}J_{\rm \nu}}{h_{\rm p}\nu}{\sigma_{\rm i}}({\nu})d\nu,
\end{equation}

\begin{equation}
\label{Eq:Cool}
{g_{\rm i}} = \int_{\nu_{\rm i}}^{\infty}\frac{4{\pi}J_{\rm \nu}}{h_{\rm p}\nu}h_{\rm p}(\nu - \nu_{\rm i}){\sigma_{\rm i}}({\nu})d\nu,
\end{equation}

\noindent
where ${\sigma_{\rm i}}({\nu})$ and $\nu_{\rm i}$ are the
photoionisation cross-section and frequency at the ionisation edge for
species $i$, respectively.  The temperature evolution is then computed
by solving:

\begin{equation}
\label{Eq:Thermal}
 \frac{dT}{dt} = \frac{(\gamma - 1){\mu}m_{\rm p}}{k_{\rm B}{\rho}}\left[G(t) - \Lambda(t, n_{\rm i})\right] - 2H(t)T + \frac{T}{\mu}\frac{d{\mu}}{dt},
\end{equation}

\noindent
where $\gamma = \frac{5}{3}$, and $G=\sum_{\rm i}n_{\rm i}g_{\rm i}$
and $\Lambda$ are the total photo-heating and cooling rates per unit
volume, respectively. We ignore any heating resulting from the growth
of density fluctuations as this is a minimal effect throughout the
short timescale of reionisation. Our code for solving
equation~(\ref{Eq:Thermal}) follows photo-ionisation and heating,
collisional ionisation, radiative cooling, Compton cooling and
adiabatic cooling for six species (H~$\rm \scriptstyle I$, H~$\rm
\scriptstyle II$, He~$\rm \scriptstyle I$, He~$\rm \scriptstyle II$,
He~$\rm \scriptstyle III$, $\rm e^{-}$).  The non-equilibrium
abundances of these species are found by additionally solving three
further differential equations coupled to equation~(\ref{Eq:Thermal})
(e.g. \citealt{Anninos1997,BoltonHaehnelt2007b}), and taking a uniform
clumping factor of $C = 3$ for all species. We use the rates compiled
in \cite{BoltonHaehnelt2007b} with the exception of the case-B
recombination and cooling rates of \cite{HuiGnedin1997} and the
photo-ionisation cross-sections of \cite{Verner1996}.  The gas
density, $\rho$, is evaluated from our linear over-density field using
a standard log-normal fit (\citealt{ColesJones1991}), which
approximates the real non-linear density for the majority of voxels.
Whilst this does not adequately reproduce the extreme over-dense and
underdense tails of the density distribution (\citealt{Becker2007}),
its use does not represent a significant source of error since the
temperature measurements we compare to are smoothed over the measured 
quasar near-zones ($\sim 5$ proper Mpc). Inaccuracies in evaluating
the thermal history of the small number of regions that fall within
these tails are therefore unimportant.

Using this prescription, we evolve the temperature of each voxel,
independent of all others from an initial temperature $T_{\rm 0} =
2.27(1 + z) {\rm K}$ and an initially neutral state. This procedure
yields a final temperature distribution over our simulation volume as
a function of redshift.

\subsection{Inclusion of quasar \HeII photo-heating}
\label{SubSec:Quasar Heating}

The final part of our model is including the effect of the quasars
themselves on the thermal state of the IGM in their vicinity.  The
ionising flux from a quasar will significantly alter the thermal state
of the surrounding IGM, as its harder, non-thermal spectrum will also
reionise He~$\rm \scriptstyle II$ (\citealt{Bolton2010}). For any 
given quasar from \cite{Bolton2011a}, placed in the simulation volume at a random 
position, we account for this heating by 
augmenting the specific intensity in each voxel within the quasar
proximity zone by an additional term

\begin{equation}
\label{Eq:Quasar Flux}
J_{{\nu},\rm q} \simeq \frac{L_{\nu}}{(4\pi R)^{2}}\exp\left[-\frac{R}{\lambda_{\rm HI}}\left(\frac{\nu}{\nu_{\rm HI}}\right)^{-3(\beta-1)}\right], \end{equation}

\noindent
where $R$ is the proper distance from the quasar and $L_{\nu,\rm
  q}=L_{\rm HI,q}(\nu/\nu_{\rm HI})^{-\alpha_{\rm q}}$ is the quasar
ionising luminosity, again taken to be a power law spectrum with index
${\alpha_{\rm q}}$.  There is considerable uncertainty concerning the
exact value of this exponent towards high redshift.  We therefore
allow it to vary over the range $1.0 \leq \alpha_{\rm q} \leq 2.0$,
which is consistent with recent indirect estimates from observations
of quasar near-zone sizes at $z > 6$ (\citealt{WyitheBolton2010}).  To
compute the quasar luminosities at the Lyman limit, we take the
absolute AB magnitudes from \cite{Carilli10}.  Finally, we fix the
extent of the region which experiences additional ionisation and
heating by each quasar to 5 proper Mpc, matching the radius within
which the near-zones were analysed in \cite{Bolton2011a}.


\section{Modelling of ionising sources}
\label{Sec:Nion Models}

We next discuss the parametrisation of ionising sources in our model,
encapsulated by the redshift-dependent parameter $N_{\rm ion}$, the
number of ionising photons entering the IGM per baryon in galaxies.
We begin by describing existing observational constraints on the IGM
ionisation state and then proceed to discuss how these are used to
calibrate $N_{\rm ion}$ and its evolution with redshift.

\subsection{Observational constraints on $N_{\rm ion}$}
\label{SubSec:Nion Constraints}

\subsubsection{The ionisation rate from the \Lya forest opacity}
\label{SubSubSec:Ionisation Rate}

\begin{table}
\caption{Photo-ionisation rate constraints from the \Lya forest
  opacity obtained by \citet{Bolton2005} and \citet{WyitheBolton2010}.
  The most likely derived value and errors at 68 per cent confidence intervals
  for the corresponding number of ionising photons entering the IGM
  per baryon in galaxies, $N_{\rm ion}$, are also given.}
\begin{center}
\begin{tabular}{c|c|c}
\hline
Redshift & $\Gamma_{\rm -12}$ & $N_{\rm ion}$ \\
\hline
4 & $0.97\pm^{0.48}_{0.33}$ & $35.3\pm^{17.5}_{12.1}$\\
\\
5 & $0.47\pm^{0.3}_{0.2}$ & $22.7\pm^{14.3}_{9.7}$\\
\\
6 & $0.18\pm^{0.18}_{0.09}$ & $13.0\pm^{13.0}_{6.5}$\\
\hline
\end{tabular}
\end{center}
\label{tab:constraints}
\end{table}

Observations of the Ly${\alpha}$ forest in quasar absorption spectra
can be used to infer the ionising background due to luminous
sources. The measured quantity is the mean transmission of
Ly${\alpha}$ flux, $\left<F\right>$, along the line of sight, or
equivalently the effective optical depth ${\tau}_{\rm eff} = -{\rm
  log}\left<F\right>$.  With knowledge of the physical properties of
the IGM, these can be converted into the background photoionisation
rate per hydrogen atom (${\Gamma}_{\rm HI} = {\Gamma}_{\rm -12} \times
10^{-12} {\rm\, s^{-1}}$).  Using full hydrodynamical simulations of
the IGM, \cite{Bolton2005} and \cite{WyitheBolton2010} have measured
$\Gamma_{\rm -12}$ from the \Lya forest opacity at $z = 4 - 6$. Their
derived values are shown in Table \ref{tab:constraints}.  These
measurements may then be compared to our own simulated ionisation
rate, which is calculated using equation~(\ref{Eq:Gamma}). When
combined with equations ~(\ref{Eq:Ndot1})--(\ref{Eq:Flux}), and
evaluated for the case of a uniformly ionised IGM at $z < 6$ this
yields constraints on $N_{\rm ion}$, which are also shown in Table
\ref{tab:constraints}.

\subsubsection{Thomson optical depth to CMB photons}
\label{SubSubSec:Optical Depth}

The electron scattering optical depth provides an integrated
constraint on the IGM ionisation state throughout the epoch of
reionisation. We use results from the latest WMAP observations,
${\tau}_{\rm CMB} = 0.088 \pm 0.015$ (\citealt{Komatsu2011}). This is
related to the ionisation history by

\begin{equation}
{\tau}_{\rm CMB} = \int_{0}^{z_{\rm CMB}}c\frac{dt}{dz}Q(z)\left(1 + f_{\rm He}(z)\right)n_{\rm H}(z){\sigma_{\rm T}}dz, 
\end{equation}
where ${\sigma_{\rm T}}$ is the Thomson scattering cross-section,
$n_{\rm H}$ is the hydrogen number density and $f_{\rm He}$ is a
correction factor due to the presence of ionised helium. In our model,
we assume that the ionisation of \HeI traces that of H$\,\rm
\scriptstyle I$, whilst \HeII is instantaneously ionised at $z = 3$.

\subsection{Parametrisation of $N_{\rm ion}$ }
\label{SubSec: Parametrising Nion}

\begin{figure*}
\begin{center}
  \includegraphics[width=0.90\textwidth]{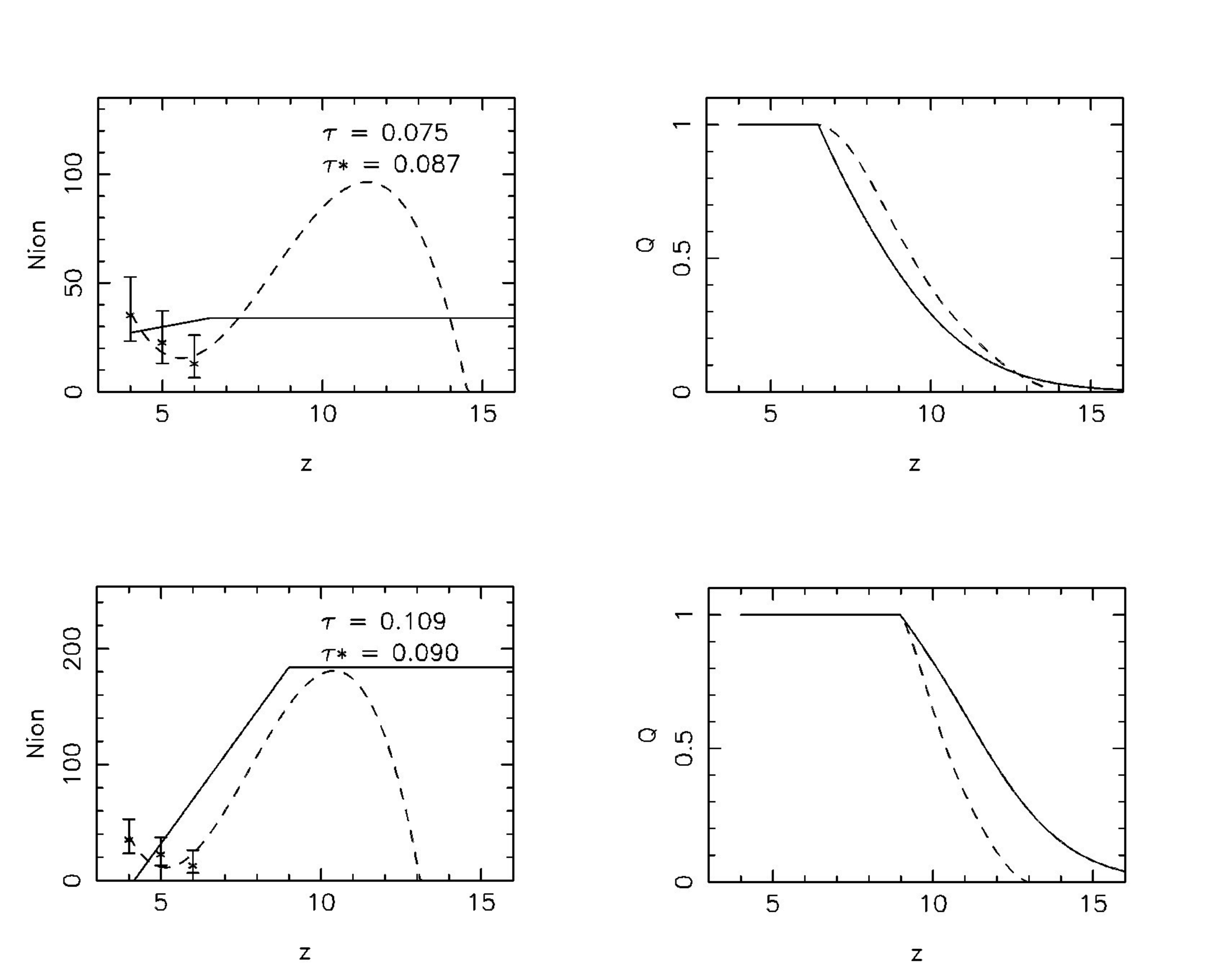}
\vspace{-0.3cm}
\caption{Evolution of the number of ionising photons per baryon in
  galaxies which escape into the IGM, $N_{\rm ion}$ (left hand
  columns), and the \HII filling factor, $Q$ (right hand columns) for
  reionisation histories with $z_{\rm r} = 6.5$ (top) and $z_{\rm r} =
  9.0$ (bottom).  Solid curves show the evolution for the linear
  $N_{\rm ion}$ model, whilst the case of the cubic $N_{\rm ion}$ is
  shown by the dashed lines. The plotted points represent constraints
  imposed by values of ${\Gamma_{\rm -12}}$ obtained from the \Lya
  forest opacity by \citet{Bolton2005} and
  \citet{WyitheBolton2010}. The inset in the left hand panels gives
  the electron scattering optical depth for the linear and cubic
  $N_{\rm ion}$ models ($\tau$ and $\tau*$, respectively).}
\label{Fig:Nion}
\end{center}
\end{figure*}

Using these observations, we may now constrain possible
parametrisations of $N_{\rm ion}$ in our model.  There is considerable
uncertainty concerning the exact shape of $N_{\rm ion}$, particularly
for $z > 6$, where there are no direct observations. Indeed, several
studies (\citealt{Pritchard2009, Mitra2011}) have found that $N_{\rm
  ion}$ is effectively unconstrained in the range $6 < z < 11$. To
explore the sensitivity of our simulated temperatures to changes in
the ionising sources, we therefore choose to employ two different
models for the redshift evolution of $N_{\rm ion}$.

\subsubsection{Linear model}
\label{SubSubSec: Constant Model}

\begin{figure*}
\begin{center}
 \includegraphics[width=1.0\textwidth]{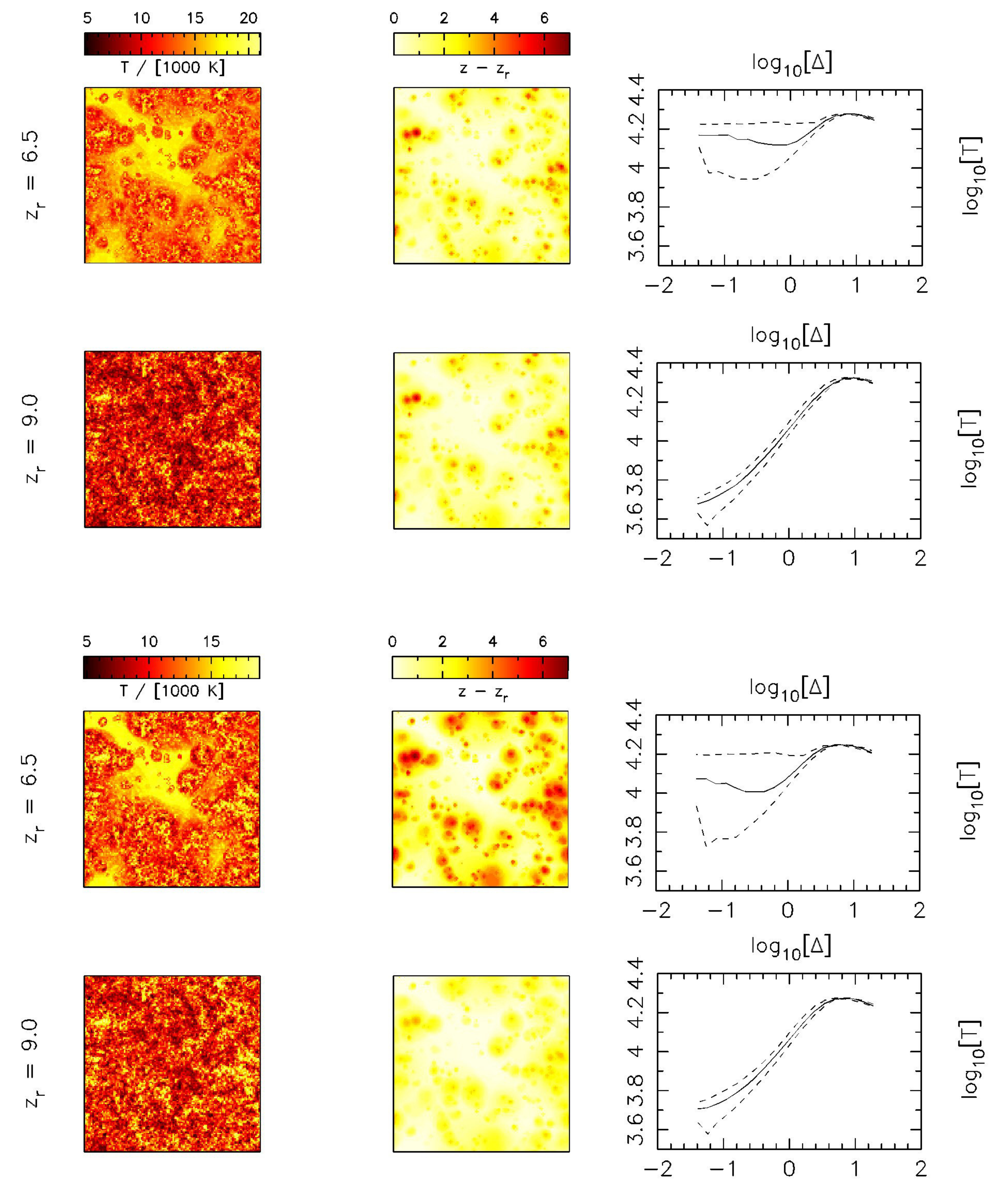}
\vspace{-0.3cm}
\caption{Maps of a two-dimensional slice through the centre of our
  $100h^{-1}$ comoving Mpc simulation box at $z = 6.0$ prior to the
  inclusion of any quasar heating. The upper six panels show results
  for the linear $N_{\rm ion}$ model while the lower set show the same
  results allowing for a cubic $N_{\rm ion}$.  In each set, the upper
  row indicates maps for reionisation occurring at $z_{\rm r} = 6.5$
  and the lower row for $z_{\rm r} = 9.0$. {\em Left panels:} The IGM
  temperature at the measurement redshift ($z = 6.0$). {\em Central
    Panels:} Maps of the ionisation redshift for each voxel relative
  to the redshift at which reionisation completes, $\Delta z = z -
  z_{\rm r}$. {\em Right panels:} The IGM temperature, T, against
  overdensity, $\Delta$, at $z = 6.0$, displaying both the median
  (solid) and 68 per cent intervals around the median (dashed).}
\label{Fig:allmaps}
\end{center}
\end{figure*}

The simplest model, and one often used in previous simulations of
reionisation (e.g. \citealt{HaimanLoeb1997,geil2008}), assumes a
single, redshift independent parametrisation of $N_{\rm ion}$.
However, this cannot be reconciled with measurements of the
photo-ionisation rate from the \Lya forest at $z\leq 6$, so we employ
a slight variant on the model. We take $N_{\rm ion}$ to be a constant
up to the reionisation redshift, which is found by solving
equation~(\ref{Eq:History}) for a universe that is fully ionised by
the desired redshift. This effectively fixes the reionisation history
for any given $z_{\rm r}$. Then, to ensure consistency with
observations of $\Gamma_{\rm -12}$, we allow $N_{\rm ion}(z)$ to
evolve linearly once reionisation is complete in order to best fit the
data points at $z < 6$. This is broadly consistent with predicted
increases in the escape fraction towards high redshift
(\citealt{Wyithe2010}).  As part of our constraint on $z_{\rm r}$ we
impose the resulting electron scattering optical depth as an
a~posteriori constraint on $z_{\rm r}$ alongside the predicted
temperatures in Section \ref{Sec:Constraints}.

\subsubsection{Cubic Model}
\label{SubSubSec: Cubic Model}

We also explore arbitrary ionisation histories that best match the
${\tau}_{\rm CMB}$ constraint, by employing a model that is
redshift-dependent prior to $z_{\rm r}$. To ensure enough freedom to
match all constraints, we choose to follow the approach of
\cite{Pritchard2009} and model $N_{\rm ion}$ as a cubic function,
allowing for a large variety of different evolutions:

\begin{equation}
\label{Eq:Nion}
N_{\rm ion}(z) = \left(N_{\rm p} - N_{\rm 0}\right)\left[4\left(\frac{z - z_{\rm 0}}{\Delta z}\right)^{3} - 3\left(\frac{z - z_{\rm 0}}{\Delta z}\right)\right] + N_{\rm 0},
\end{equation}

\noindent
where we have parametrised by the midpoint $\left(z_{\rm 0}, N_{\rm
  0}\right)$, and the peak $\left(z_{\rm 0} + \frac{\Delta z}{2},
N_{\rm p}\right)$. For each reionisation redshift, we fit this curve
to the measured electron scattering optical depth and the ionisation
rate at $z = 4 - 6$. As in \cite{Pritchard2009}, we impose a prior on
the comoving emissivity of $\dot{N_{\rm \gamma}} < 10^{51} {\rm\,
  s^{-1}} {\rm Mpc^{-3}}$ to disallow unphysically large ionising
backgrounds. We stress that this model is not physically motivated and
clearly overfits the limited data we use. However, as we do not seek
to exactly find $N_{\rm ion}$ but only to view the effect of different
ionisation histories on simulated IGM temperatures, this does not
significantly impact upon our results.

\subsubsection{Optimal $N_{\rm ion}$ models}

Finally, the best fit evolution of $N_{\rm ion}$ and the filling factor
$Q$ for the cubic and linear models are shown in Figure \ref{Fig:Nion}
for $z_{\rm r} = 6.5$ and $z_{\rm r} = 9.0$.  For both cases, the
behaviour of the emissivity matches quite well to results derived
elsewhere (\citealt{ChoudhuryFerrara2006, Pritchard2009, Mitra2011}),
although note that for early reionisation in the cubic model, star
formation switches on later and the filling factor develops more
quickly in order to match the electron scattering constraint.
Importantly, the two models do provide very different ionisation
histories at each $z_{\rm r}$ and therefore should allow us to probe
the sensitivity of our temperature predictions to realistic variations
in the ionising background.


\section{Results of the semi-numerical model}
\label{Sec:Results}

\subsection{The IGM thermal state following hydrogen reionisation}
\label{SubSec:Thermal Results}

We now examine the variation of IGM temperature with
reionisation redshift using the semi-numerical model described in the
previous two sections.  For any given $z_{\rm r}$, and for each of the
two different parametrisations of $N_{\rm ion}$, we compute an optimal
$N_{\rm ion}$ evolution that ensures $Q = 1.0$ by $z_{\rm r}$. The
resulting ionising background then allows us to calculate the
temperature field over the redshift range $5.5 < z < 6.5$, when the
quasars analysed in \cite{Bolton2011a} are expected to have had their
most recent optically bright phase.

The IGM temperatures predicted by our semi-numerical model at $z =
6.0$ are summarised in Figure~\ref{Fig:allmaps}.  The top two rows
display the results using our linear $N_{\rm ion}$ model, for
reionisation completing by $z_{\rm r} = 6.5$ (first row) and $z_{\rm
  r} = 9.0$ (second row).  The left hand panels display temperature
maps through the centre of the $100h^{-1}$ Mpc simulation volume.  The
corresponding central panels show maps of the voxel ionisation
redshifts relative to the redshift at which reionisation completes
(i.e. $\Delta z = z - z_{\rm r}$).  We note that the structure of
ionised bubbles at fixed $\Delta z$ is very similar for the early and
late models.  In the case of late reionisation, the temperature mimics
this ionisation structure, with cooler regions centred on the small
\HII bubbles that were ionised first, and have thus had more time to
cool. However, regions of the IGM ionised during the overlap phase
retain a clear imprint of heating.  The situation is very different
for early reionisation, where the temperature field shows little
correlation with the reionisation topology. Instead, the temperature
largely reflects the random Gaussian structure of the density field,
indicating that even with reionisation occurring at $z_{\rm r} = 9.0$,
the temperature has already settled to an almost asymptotic state by
$z = 6.0$.

This asymptotic limit is also clear from the temperature-density
relation (right hand panels of Figure~\ref{Fig:allmaps}). For $z_{\rm
  r} = 9.0$, the relation shows an almost power law dependence with
$\gamma - 1 \approx 0.5$ for $T = T_{\rm 0}\Delta^{\gamma - 1}$,
characteristic of an asymptotic IGM thermal state which is set by the
competition between photo-heating and adiabatic cooling
\citep{HuiGnedin1997}.  However, in the case of late reionisation, the
figure shows much wider 68 per cent confidence limits (dashed lines) below
the highest densities, with an inverted power-law slope of $\gamma - 1
\approx -0.2$ for low densities ($\Delta < 1.0$). The scatter and
anti-correlation between temperature and density, which arises due to
the fact that lower density regions tend to be ionised last allowing
them less time to cool, is in qualitative agreement with results from
previous numerical and analytical studies
(\citealt{Bolton2004,Trac2008,FurlanettoOh2009}).

The same set of results can also be seen for the cubic model of
$N_{\rm ion}$ in the two lower rows in Figure~\ref{Fig:allmaps}. In
this case, the growth of ionised bubbles is significantly different at
constant $\Delta z = z-z_{\rm r}$ for the two reionisation
redshifts. As described in Section~\ref{SubSec: Parametrising Nion},
late reionisation ($z_{\rm r} = 6.5$) is a more gradual process in
this model, with ionisation times spread more evenly around the \HII
bubbles. This has the effect of spreading out the IGM temperatures, so
that regions ionised early tend to be cooler. Similarly, the scatter
in the temperature-density relation is increased. On the other hand,
for $z_{\rm r} = 9.0$, reionisation is far more rapid, effectively
occurring over a redshift interval of ${\Delta}z = 3.0$. Despite this
difference, the temperature field evolves, by $z = 6.0$, to the same
asymptotic state as for the linear $N_{\rm ion}$ model (with a
slightly reduced scatter in the temperature-density relation due to
the more abbreviated differences in ionisation time). This reaffirms
the conclusion that the thermal memory of hydrogen reionisation for
$z_{\rm r} = 9.0$ is largely erased by $z = 6.0$.

\begin{figure}
\begin{center}
\includegraphics[width=0.45\textwidth]{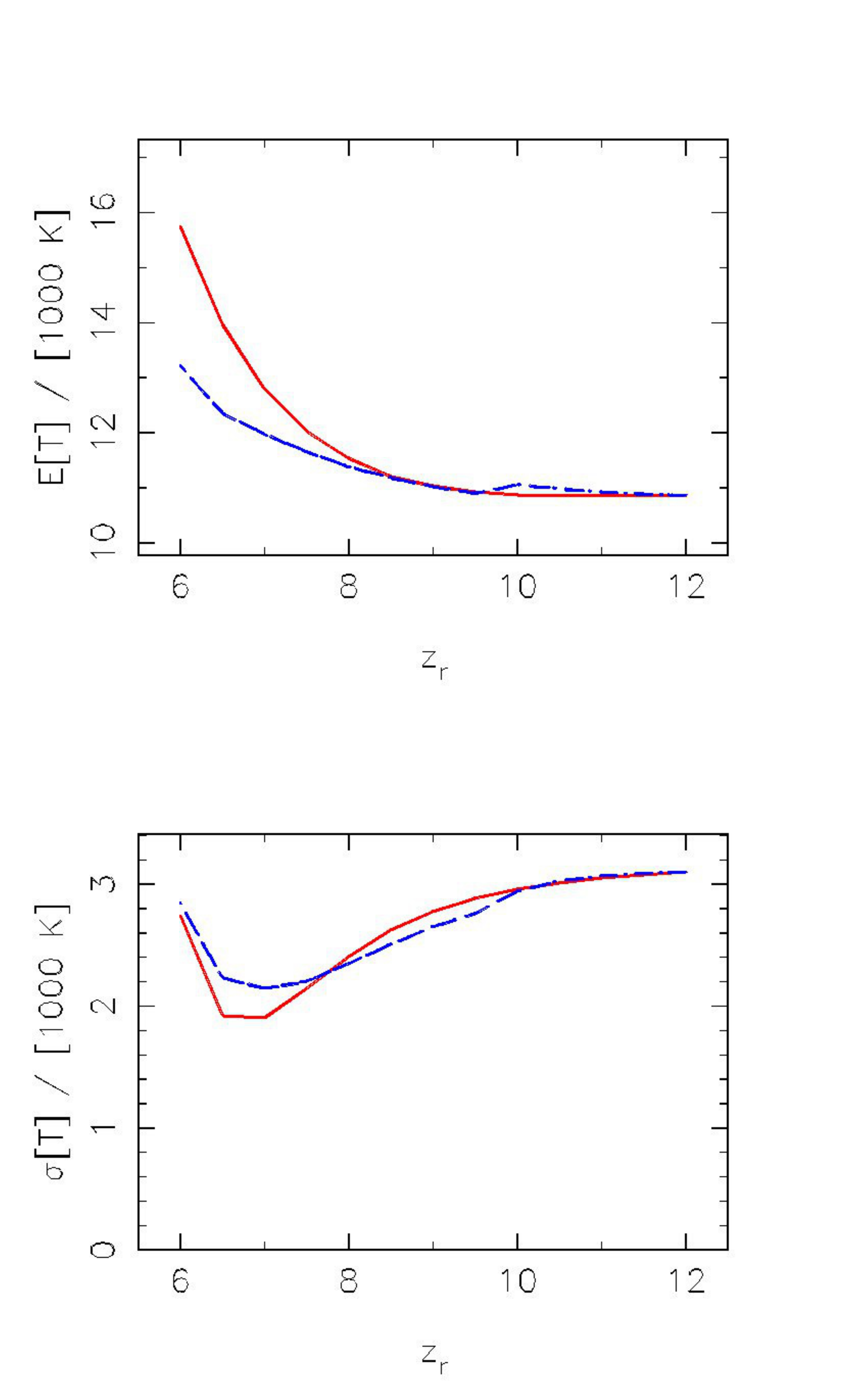}
\vspace{-0.3cm}
\caption{Mean (top panel) and standard deviation (bottom panel) of the
  simulated temperature distributions at $z = 6.0$ prior to the
  inclusion of quasar heating.  Note the results are obtained from the
  entire $100h^{-1}$ comoving Mpc simulation box, and the mean is
  therefore volume averaged over all densities.  We show results for
  two different cases; the linear $N_{\rm ion}$ model (solid red
  curves) and the cubic model allowing for an evolving $N_{\rm ion}$
  (dashed blue curves).}
\label{Fig:Tvszplot}
\end{center}
\end{figure}

To see this more clearly, in Figure \ref{Fig:Tvszplot} we consider how 
the mean and standard deviation of the temperature distribution at 
$z = 6.0$ evolve as a function of reionisation redshift. We note that the 
mean referred to here is not directly comparable to the measurements of 
\cite{Bolton2011a} as it is averaged over the entire density field in the 
simulation volume and is not simply the temperature at mean density.  The
results are shown for both the linear (solid red) and cubic (dashed
blue) $N_{\rm ion}$ models.  The mean temperature predicted by the two
models only varies by several hundred degrees Kelvin above $z_{\rm r}
\approx 9.0$, and an asymptotic state is fully reached for $z_{\rm r}
\approx 10.5 - 11.0$. However, for later reionisation redshifts the
temperature distributions vary quite rapidly, and for the linear
$N_{\rm ion}$ model at $z_{\rm r} = 6.0$ the IGM is $\sim 5000\rm\,K$
hotter than the asymptotic value. This result is slightly more muted
for the cubic model, as the elongated reionisation history means that
large regions of the IGM have longer to cool, thereby reducing the
impact of heating on the mean temperature.  However, even allowing for
this, unit changes in the reionisation redshift correspond to
temperature variations of up to $1500\rm\, K$ by $z = 6.0$.  As this
is comparable to the standard deviation of the distributions, changes
of ${\Delta}{z_{\rm r}} \approx 1.0$ should be distinguishable on the
basis of volume averaged temperature measurements for $z_{\rm r} <
9.0$, but will require very precise data.

Finally, note that the standard deviation exhibits a minimum at
$z_{\rm r} \sim 7$ for both models, but increases to both lower and
higher redshift.  At $z_{\rm r}<7$, this scatter is driven by the
large variation in temperatures resulting from recent, inhomogeneous
reionisation.  In contrast, the scatter at $z_{\rm r}>7$ is instead
due to the steepening correlation between temperature and density
which establishes itself following reionisation.  Both models
asymptotically approach a value of $\sigma(T)\sim 3000\rm\,K$ for
$z_{\rm r} \ga 10$.

\subsection{The temperature in quasar near-zones}
\label{Quasar Heating}

\label{SubSubSec: Comparisons}

\begin{figure}
\begin{center}
  \includegraphics[width=0.45\textwidth]{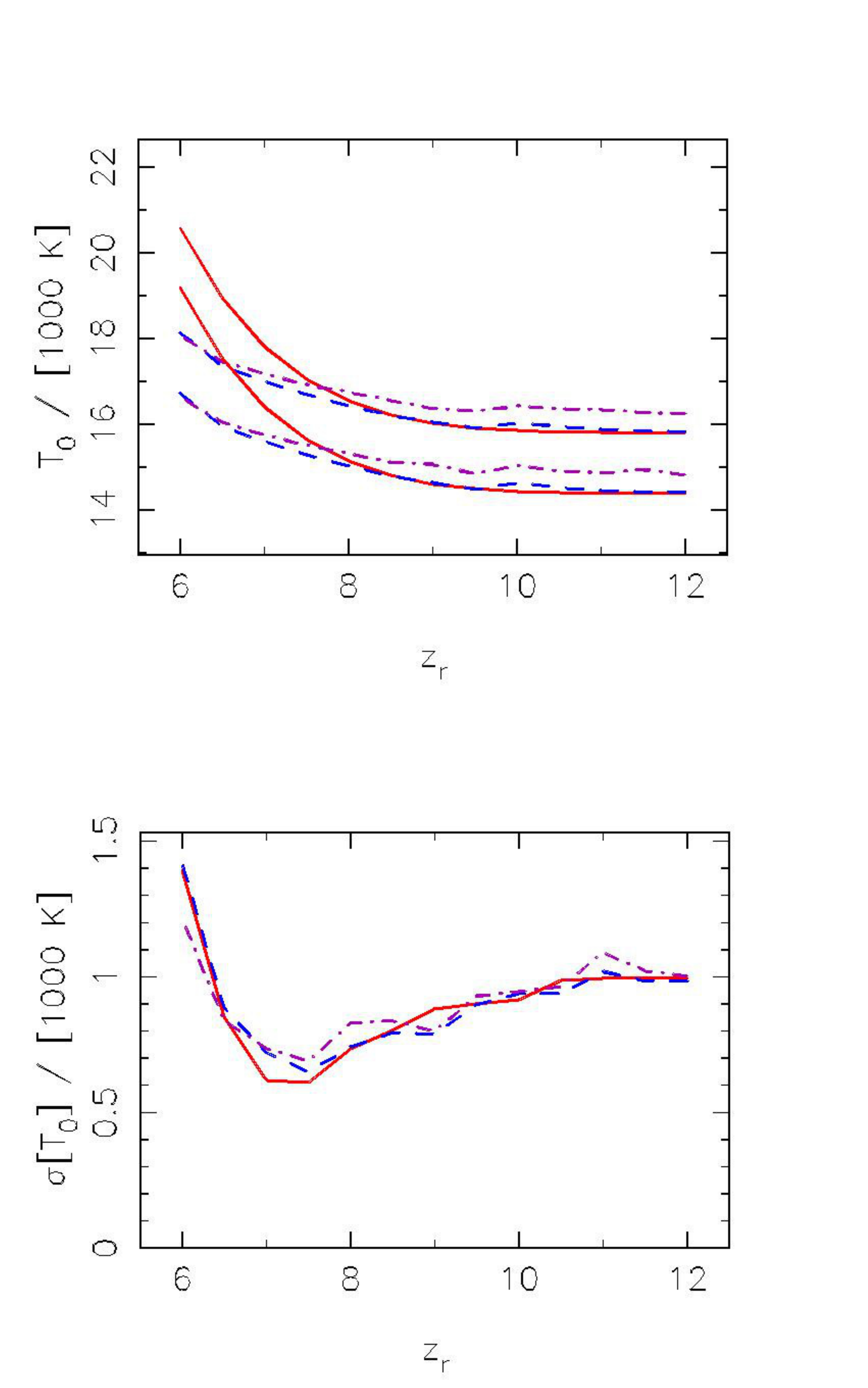}
\vspace{-0.3cm}
\caption{Mean and standard deviation of the simulated distributions over
  temperature at mean density, $p(T_{\rm 0} | z_{\rm r})$, in a quasar
  proximity zone at $z = 6.02$. {\em Top:} Simulated mean for three
  different cases, the linear $N_{\rm ion}$ model (solid lines), the
  cubic model (dashed lines), and the cubic model accounting for the
  density bias of quasars (dot-dashed lines).  We show simulated
  temperatures for $\alpha_{\rm q} = 2.0$, which yield the lower
  curves and $\alpha_{\rm q} = 1.0$ giving the higher set of
  curves. {\em Bottom:} Standard deviations for the aforementioned
  models. In this case we only show results for $\alpha_{\rm q} =
  1.0$, however standard deviations are similar regardless of quasar
  spectral index.}
\label{Fig:Tvszplotq}
\end{center}
\end{figure}

\cite{Bolton2011a} used the Doppler parameters of \HI
Ly$\alpha$ absorption lines in the near-zones of 7 quasars at $5.8 < z
< 6.4$ to measure the IGM temperature at mean density, $T_{0}$, within
$\sim 5$ proper Mpc of these sources.  In order to compare these
observations to our model, we must therefore also include the effect
of \HeII photo-heating by the quasars themselves on the IGM
temperature.  

We calculate this heating as follows.  At any given voxel in our
simulation volume, we switch on a quasar $t_{\rm q} = 1.5 \times
10^{7}$ yrs prior to the redshift at which the quasar is observed, and
use the procedure outlined in Section \ref{SubSec:Quasar Heating} to
calculate the surrounding IGM temperature.  We then draw 200 random
lines-of-sight within a $\sim 5$ proper Mpc region around the quasar,
and for each one we measure the temperature-density relation for all
voxels. This relation may then be used to infer a temperature at mean
density averaged over all 200 lines-of-sight. Repeating this procedure
for ``quasars'' centred on all voxels in the simulation box then gives
a distribution for the temperatures around a single observed quasar.

In Figure \ref{Fig:Tvszplotq} we show the mean and standard deviation
of the simulated temperature distributions {\it at mean density} for a
quasar observed at $z = 6.02$.  The results for the mean temperature
are displayed for both hard ($\alpha_{\rm q} = 1.0$, upper curves) and
soft ($\alpha_{\rm q} = 2.0$, lower curves) quasar spectral indices.
The red solid and blue dashed curves again correspond to the linear
and cubic $N_{\rm ion}$ models. These retain all the features of the
temperature distributions prior to quasar heating, including the
asymptotic behaviour of the mean at $z_{\rm r} \ga 9.0$, and the
minimum in the scatter at $z_{\rm r} \simeq 7.0$.  Note, however, the
standard deviation tends to be slightly smaller, particularly at large
$z_{\rm r}$.  This is due to the fact that we are now only considering
the average temperature at a single (mean) density.

We find that the maximum difference in the average temperature at mean
density for the cubic $N_{\rm ion}$ model is only $\sim 2\,500 {\rm
  K}$ for early and late reionisation.  Given that the 68 per cent
measurement uncertainties on the temperature around individual quasars
obtained by \cite{Bolton2011a} are $\sim 3\,000$--$5\,000 {\rm K}$, it
is unlikely that a strong constraint can be obtained from any single
line-of-sight observation.  However, the combination of all 7 quasars
analysed by \cite{Bolton2011a} reduces the 68 per cent confidence
interval on the mean temperature of the quasar near-zones to $\sim
1\,500 {\rm K}$.  Therefore, all seven lines-of-sight will be able, in
principle, to distinguish between early and late reionisation
scenarios.

It is important to note that the rather uncertain quasar spectral
index also introduces additional uncertainties into the interpretation
of temperature.  For example, in the cubic model the temperatures
obtained for late reionisation combined with a soft quasar spectrum
are almost identical to those obtained for early reionisation and a
much harder quasar spectrum.  Unfortunately, there are no direct
observations that constrain high-redshift quasar spectral indices,
although a recent study of quasar near-zone sizes at $z\simeq 6$
demonstrated that the range $1.0 \leq \alpha_{\rm q} \leq 2.0$ is
consistent with the data (\citealt{WyitheBolton2010}).  This is in
agreement with direct observations at lower redshift ({\it e.g.}
\citealt{Telfer2002}), but in practice the quasar spectral index will
remain an uncertainty in our analysis.

\subsection{Quasar Density Bias}
\label{SubSubSec:Galaxy Bias}

When calculating the measured temperature distribution expected around
the quasars, we have thus far assumed a random distribution of quasar
positions in our simulation volume.  However, quasars and their host
galaxies tend to trace high density regions owing to the clustering
bias of massive haloes.  Before calculating the constraints on $z_{\rm
  r}$, we therefore also model the quasar density bias. We weight the
temperature distribution calculated in Section~\ref{SubSubSec: Comparisons} by
the likelihood $\mathcal{L}_{\rm g}(\delta)$ of observing a quasar at
that location

\begin{equation}
p(T | w)  = \sum_{all voxels}^{}\delta(T(x) - T)\mathcal{L}_{\rm g}(\delta(x)).
\end{equation}
The likelihood of observing a galaxy, and a corresponding resident
quasar, is then estimated from the \cite{ShethTormen2002} mass
function as

\begin{equation}
\label{Eq:LH}
\mathcal{L}_{\rm g}(\delta) = \frac{(1+\delta)\nu(1+\nu^{-2p})
e^{-a\nu^2/2}}{\bar{\nu}(1+\bar{\nu}^{-2p})e^{-a\bar{\nu}^2/2}},
\end{equation}
where $\nu=(\delta_{\rm c}-\delta)/[\sigma(M)]$, $\bar{\nu} = \delta_{\rm c}/[\sigma(M)]$, 
and $a=0.707$, $p=0.3$ are constants.

The results including this density bias within the cubic $N_{\rm ion}$
model are shown by the dot-dashed curves in Figure
\ref{Fig:Tvszplotq}.  There is a slight flattening of the temperature
differences between different reionisation redshifts.  This is due to
the relatively early reionisation time of the high density regions
where quasars are found, which allows them more time to cool to their
asymptotic thermal state. At large values of $z_{\rm r}$ the
temperature is also systematically higher.  This is because the higher
recombination rate in higher density regions results in slightly
larger thermal energy input from photo-heating.  This increased mean
temperature makes it slightly more difficult for measurements of
$T_{\rm 0}$ to differentiate between reionisation redshifts. However,
this is a minimal effect which will not strongly affect our
constraints on $z_{\rm r}$.


\section{Constraints on the reionisation history}
\label{Sec:Constraints}

In the remainder of the paper we turn to calculating the constraints
that recent and possible future IGM temperature measurements at
$z\simeq 6$ impose on the hydrogen reionisation redshift.  Aside from
a number of fixed assumptions including the clumping factor $C$, star
formation mass thresholds $M_{\rm min}, M_{\rm ion}$, the stellar
spectral index $\alpha$, and the evolution of $N_{\rm ion}$, there are
two critical free parameters in our model, $z_{\rm r}$ and
$\alpha_{\rm q}$.  For any combination of these two free parameters,
we compute mock observations for each $z_{\rm r}$ and $\alpha_{\rm q}$
for the $i^{\rm th}$ near-zone temperature measurement, $q^{\rm
  i}(z_{\rm r}, \alpha_{\rm q}^{\rm i})$. Using Bayes' Theorem, we
then infer the likelihood for any set of 7 independent sightlines
$\{q^{\rm i}\}_{i = 1}^{7}$ (corresponding to the 7 measured
temperatures from \citealt{Bolton2011a}):

\begin{equation}
\label{Eq:Bayes}
p(\{q^{\rm i}\}_{i = 1}^{7} | {\bf D}, M)=\frac{p({\bf D} | \{q^{\rm i}\}_{i = 1}^{7}, M) p (\{q^{\rm i}\}_{i = 1}^{7} | M)}{p({\bf D} | M)},
\end{equation}
where ${\bf D}$ is the combination of observational constraints
obtained for $\tau_{\rm CMB}$, the ionising background and the quasar
near-zone temperature measurements, $M$ is one of our models for
reionisation, and $p({\bf D} | M)$ is a normalisation constant. The
likelihood $p({\bf D} |\{q^{\rm i}\}_{i = 1}^{7}, M)$ can be easily
found from the measured temperature distributions in Figure
\ref{Fig:Tvszplot} and the measured errors on $\tau_{\rm CMB}$ (which
are taken to be Gaussian) and $\Gamma_{-12}$. The prior probability of
choosing any set of sight lines $p(\{q^{\rm i}\}_{i = 1}^{7} | M)$ is
calculated from the density bias likelihood in equation~(\ref{Eq:LH})
and the assumption that prior information on our free parameters is
uniform over the ranges of interest, $1.0 \leq \alpha_{\rm q} \leq
2.0$ and $6.0 \leq z_{\rm r} \leq 12.0$.  Redshifts outside this range
are not considered as they are effectively ruled out at the lower end
by the Ly$\alpha$ forest data (\citealt{Fan2006}) and at the upper end
by the measurement of $\tau_{\rm CMB}$ (\citealt{Komatsu2011}).

Using equation~(\ref{Eq:Bayes}), we may then find absolute bounds on
the reionisation redshift by summing over sets of possible sightlines

\begin{equation}
\label{Eq:Marginalise}
p(z_{\rm r} | {\bf D}, M)=\sum_{\{q^{\rm i}\}} p(\{q^{\rm i}\}_{i = 1}^{7} | {\bf D}, M) \delta[z_{\rm r}(\{q^{\rm i}\}_{i = 1}^{7} | M) - z_{\rm r}],
\end{equation}
where the Kronecker delta $\delta$ arises since all seven sightlines are drawn 
from a single simulation box and correspond to one value of $z_{\rm r}$.  We note,
however, that the same does not hold true for the quasar spectral
indices, as each of the individual quasars may have different
intrinsic spectra. Therefore, we only obtain constraints for each
quasar spectral index individually:

\begin{equation}
\label{Eq:Marginalisealpha}
p(\alpha_{\rm q}^{\rm i} | {\bf D}, M)=\sum_{\{q^{\rm i}\}} p(\{q^{\rm i}\}_{i = 1}^{7} | {\bf D}, M) \delta[\alpha_{\rm q}^{\rm i}(\{q^{\rm i}\}_{i = 1}^{7} | M) - \alpha_{\rm q}^{\rm i}]
\end{equation}

\noindent
Equations~(\ref{Eq:Marginalise}) and (\ref{Eq:Marginalisealpha})
produce model dependent estimates for our parameters of interest. We
note that there are several uncertain aspects of our modelling, which
can lead to significant variations in sightline temperature,
particularly for late reionisation (Figure~\ref{Fig:Tvszplotq}).

To both compare the relative probability of our different models, and
also obtain a less model dependent estimate of $z_{\rm r}$ and
$\alpha_{\rm q}^{\rm i}$, we use a Bayesian model comparison to
calculate the posterior probability for model $M_{\rm i}$ given the
data $p(M_{\rm i} | {\bf D})$

\begin{equation}
\label{Eq:BMC}
p(M_{\rm i} | {\bf D}) = \frac{p({\bf D} | M_{\rm i})p(M_{\rm i})}{p({\bf D})}.
\end{equation}
Here $p({\bf D})$ is again a normalisation constant and we
assume a priori that all models are equally likely so that $p(M_{\rm
  i})$ is uniform. Model averaging can then be used to infer
probabilities of free parameters across all models

\begin{equation}
\label{Eq:BMA}
p(z_{\rm r} | {\bf D}) = \sum_{i}p(z_{\rm r} | {\bf D}, M_{\rm i})p(M_{\rm i} | {\bf D}),
\end{equation}
which provides a more general set of constraints on $z_{\rm r}$ and
$\alpha_{\rm q}^{\rm i}$. These constraints are not completely model
independent as the set of models we consider is not complete. However,
as our models bracket most of the physical uncertainties, they should
be reliable.

\subsection{Reionisation constraints from measured near-zone temperatures}
\label{SubSec:Current}

\begin{figure*}
\begin{center}
\includegraphics[width=1.0\textwidth]{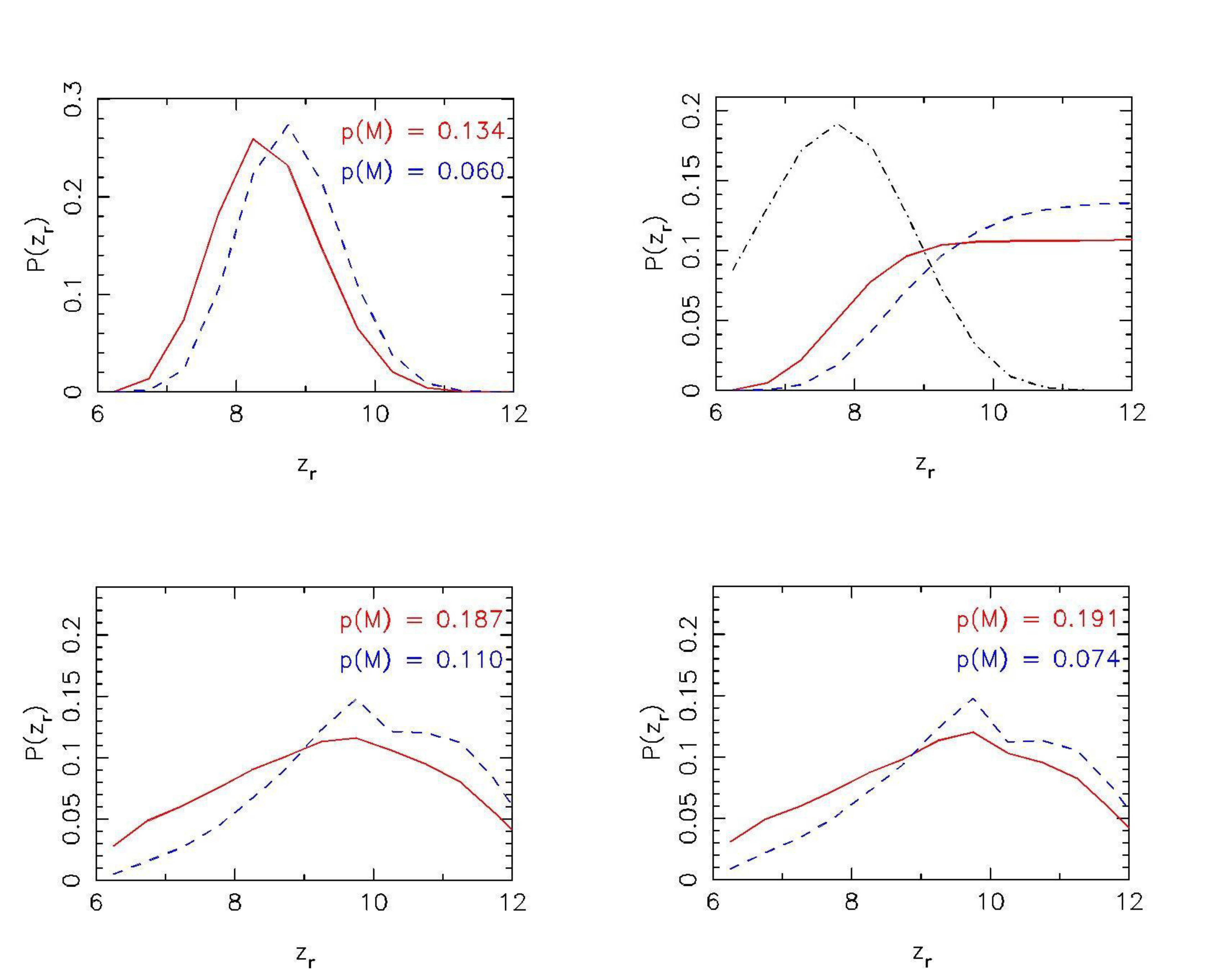}
\vspace{-0.3cm}
\caption{Constraints on reionisation redshift based on measured values
  of $T_{\rm 0}$, $\tau_{\rm CMB}$ and $\Gamma_{\rm -12}$ for
  different models of reionisation. In each case two probability
  distributions are shown, representing constraints from the fiducial
  near-zone temperature measurements of \citet{Bolton2011a} (red solid
  curves) and allowing for a larger Jeans scale (blue dashed
  curves). The relative probabilities of each model are shown in the
  top corner of each plot. {\em Upper Left:} Constraints from $T_{\rm
    0}$ and $\tau_{\rm CMB}$ and the linear model for $N_{\rm
    ion}$. {\em Upper Right:} Constraints obtained from the near-zone
  temperatures and $N_{\rm ion}$ alone. For comparison, the constraint
  from $\tau_{\rm CMB}$ and $N_{\rm ion}$ alone is shown in black
  dot-dashed. {\em Lower Left:} The probability distributions for the
  cubic $N_{\rm ion}$ model.  {\em Lower Right:} Distributions for the
  cubic model but now accounting for the bias of quasars to regions of
  high density.}
\label{Fig:zrconstraints}
\end{center}
\end{figure*}

In the upper panels of Figure \ref{Fig:zrconstraints}, we present the
resulting probability distribution functions over the parameter
$z_{\rm r}$ for the linear $N_{\rm ion}$ model.  The distributions are
computed for the two different sets of measurements presented by
\cite{Bolton2011a}, derived from their fiducial model (red solid
curves) and a model with additional pressure (Jeans) smoothing in the
IGM (blue dashed curves).  The latter effectively lowers the near-zone
temperature constraints by $\sim 3000\rm\,K$.  The upper left panel
displays the $p(z_{\rm r})$ based on the measured values of $T_{0}$,
$\tau_{\rm CMB}$ and $\Gamma_{-12}$.  For comparison, in the upper
right panel we also plot the individual distributions obtained by
alternately considering constraints from either $\tau_{\rm CMB}$ or 
$T_{\rm 0}$ alone combined with our assumed $N_{\rm ion}$ evolution.

As expected, the measurement of $T_{\rm 0}$ alone does not
differentiate between early reionisation redshifts, as for all
scenarios the IGM has reached an asymptotic temperature. However, the
temperature data do provide constraining power for low $z_{\rm r}$.
We find that irrespective of which set of temperature measurements
from \cite{Bolton2011a} is considered, very late reionisation is ruled
out to high confidence. This is because the $5\,000 {\rm K}$ increase
in IGM temperature for $z_{\rm r} < 6.5$, pushes the mean quasar
temperature well above the observations.

In the lower panels of Figure \ref{Fig:zrconstraints} we show
corresponding constraints for the cubic model excluding (lower left
panel) and including (lower right panel) the effect of quasar density
bias.  In this model the observed value of $\tau_{\rm CMB}$ is used
in defining the parameters governing $N_{\rm ion}$. It is therefore a
much weaker a-posteriori constraint and we can no longer rule out 
reionisation at high redshift.  Similarly for low $z_{\rm r}$, the
smaller temperature differences between reionisation scenarios means
that late reionisation cannot be excluded with the same
confidence. Furthermore, the relatively tight constraints obtained
when considering the temperatures derived assuming a larger Jeans
scale are probably not robust, as the constraints from the fiducial
measurements are favoured by a factor of around 3 over these.

\begin{figure}
\begin{center}
  \includegraphics[width=0.45\textwidth]{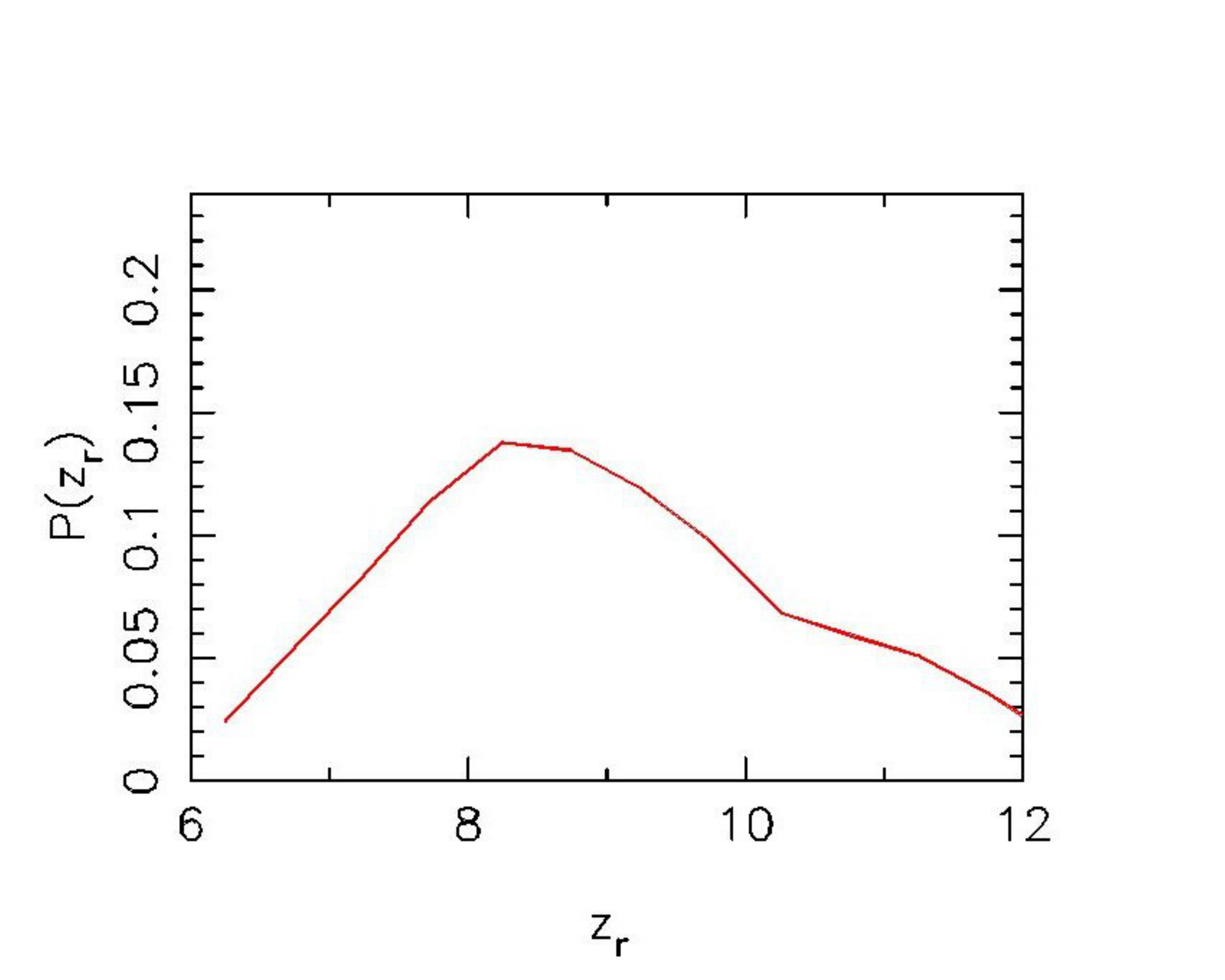}
\vspace{-0.3cm}
\caption{Final constraint on the reionisation redshift based on
  measured values of $T_{\rm 0}$, $\tau_{\rm CMB}$ and $\Gamma_{\rm
    -12}$ when averaged over all our different models of
  reionisation. This includes different models for $N_{\rm ion}$, the
  bias of quasars to high density regions and the Jeans pressure
  smoothing scale. Model averaging significantly weakens constraints
  as it adds uncertainties to the temperature, but still rules out
  late reionisation to $z_{\rm r} > 7.9$ $(6.5)$ at 68 per cent
  confidence and 95 per cent confidence respectively.}
\label{Fig:Constraintsaverage}
\end{center}
\end{figure}

\begin{figure*}
\begin{center}
\includegraphics[width=1.0\textwidth]{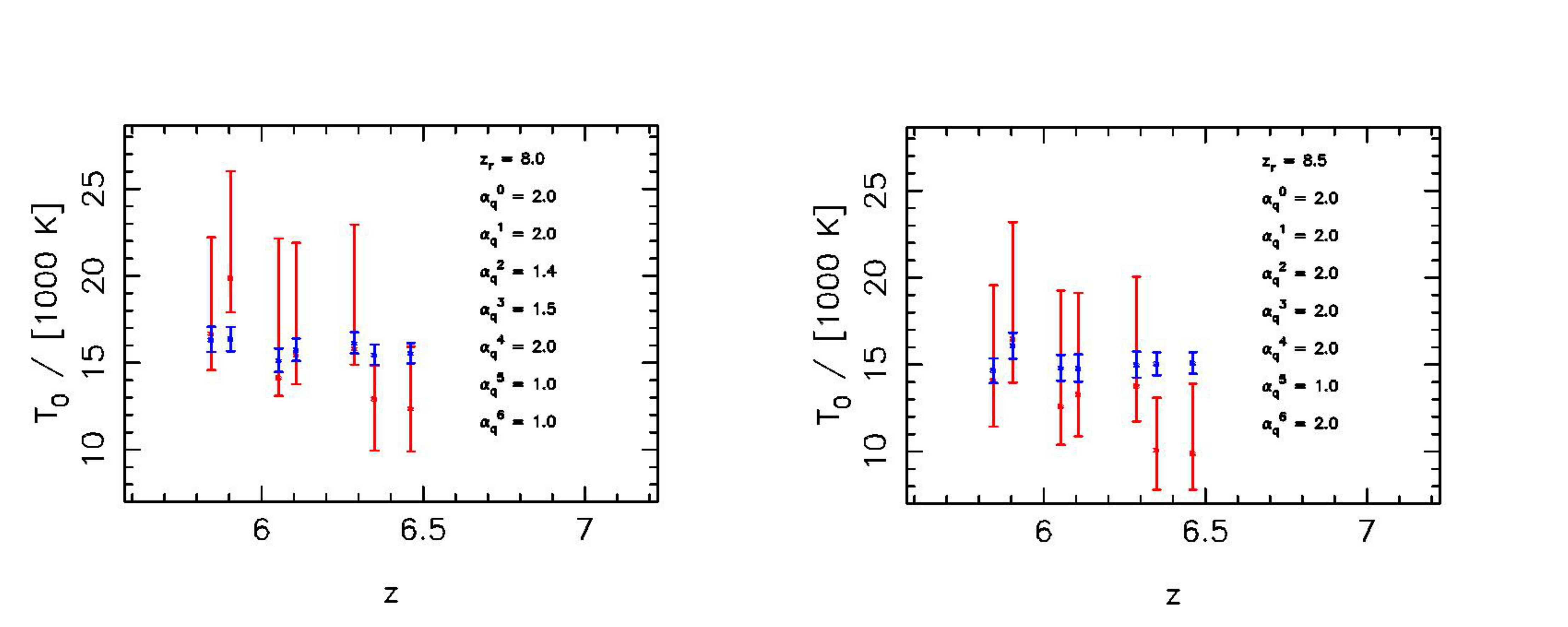}
\vspace{-0.3cm}
\caption{Model averaged quasar temperature distributions at the
  maximum a-posteriori estimate of reionisation redshift. We show both
  the model distributions (blue) and the observed sightlines (red)
  with 68 per cent confidence error bars. In both cases, we also
  include the parameter estimates of $z_{\rm r}$ and $\alpha_{\rm
    q}^{0}$ (rightmost point) through $\alpha_{\rm q}^{6}$ (leftmost
  point). For clarity, we omit redshift ranges and offset observations
  3 and 5 by $\Delta z = 0.1$. {\em Left:} Results obtained for the
  fiducial near-zone temperature measurements. {\em Right:} Results
  for the measurements with additional Jeans smoothing.}
\label{Fig:MAPestimates}
\end{center}
\end{figure*}

The final model averaged distribution is shown in
Figure~\ref{Fig:Constraintsaverage}, which corresponds to a mix of the
fiducial results from the linear and cubic models. Even though this
largely ignores the models with the tightest constraints, we may still
conclude that $z_{\rm r} > 7.9$ $(6.5)$ to 68 per cent (95 per cent)
confidence.  As a sanity check, we also verify that the optimal
temperatures predicted by our reionisation models match the
measurements quite well. {Figure \ref{Fig:MAPestimates} shows the
predicted sightline temperature distributions for the maximum
a-posteriori estimates of $z_{\rm r}$ and $\alpha_{\rm q}^{\rm i}$
compared to the observed quasar temperatures. We find that although
two thirds of the modelled points lie within the 68 per cent
confidence intervals of the observations, the model temperatures of
quasars 0 and 1 (following the numbering in Figure
\ref{Fig:MAPestimates}) appear to be systematically too high. However,
for the fiducial \cite{Bolton2011a} measurements this does not provide
a strong constraint ruling out our model, since the observed
temperature distributions are both positively skewed and broader than
Gaussian. Therefore, although our model predictions lie on the edge of
the 68 per cent confidence interval, they are only about a factor of
1.5 less probable than the peak. The same does not hold true for the
second set of measurements which account for additional Jeans
smoothing.  These are ruled out with some significance, which may be
indicate that the (conservative) systematic error due to additional
Jeans smoothing adopted by \cite{Bolton2011a} may have been slightly
overestimated.  The only real outlier then comes from quasar 5 in the
fiducial measurements, which is unsurprising since this provides a
significantly higher measurement of proximity-zone temperature.  This
may indicate that a quasar spectral index which is harder than the
prior we have assumed (i.e. $\alpha_{\rm q}<1$) may be required to
reconcile this measurement with the simulations.}

It is interesting to contrast this result with recent evidence for a
decline in the fraction of Lyman break galaxies which exhibit strong
\Lya emission from $z=5$--$7$
\citep{Ono2011,Pentericci2011,Schenker2011}.  Assuming the \Lya
emission escape fraction does not evolve rapidly, this decline may be
attributed to a corresponding decrease in the ionised hydrogen
fraction to $Q \approx 0.5$ by $z = 7$, implying reionisation finished
late.  In addition, analyses of the size and shape of the near-zone
observed in the spectrum of the quasar ULAS J1120$+$0641 are
consistent with $Q \leq 0.9$ at $z\simeq 7.1$
(\citealt{Mortlock2011,Bolton2011b}), assuming the red damping wing is
not due to a proximate high column density absorber.  Large neutral
fractions of $\sim 50$ per cent at $z\sim 7$ may be difficult to
reconcile with our temperature measurements, which indicate $z_{\rm r}
> 7.9$ $(6.5)$ to 68 per cent (95 per cent).  Such large neutral
fractions could perhaps be reconciled with our constraint if the
average quasar spectral index is much softer than we have assumed,
resulting in lower temperatures for a given $z_{\rm r}$ due to less
\HeII photo-heating.  On the other hand we note that other possible
sources of {\it hydrogen} reionisation, such as population-III stars
or mini-quasars, have harder spectra than we consider here.  These
would therefore tend to increase the heating immediately following
reionisation.  This would then drive IGM temperatures too high to
match the \cite{Bolton2011a} observations assuming reionisation
completed late.  Consequently, sources which have harder spectra than
we have assumed would strengthen rather than weaken our conclusion
that late reionisation is unlikely.

Similar considerations also apply to uncertainties in the clumping
factor $C$, briefly mentioned in Section \ref{SubSec:Ionisation
  Model}. This is a highly uncertain parameter, whose impact we did
not fully model but instead simply assumed to take on a uniform value
of $C = 3$.  The clumping factor has an impact on temperature
evolution, since increases in clumping factor tend to increase overall
recombination, implying more photons are needed to achieve
reionisation. This results in a corresponding increase in IGM heating
on the order of $\sim 1\,000 \rm K$ for $C=5$. However, we note that
our choice of clumping factor lies at the low end of possible values
derived from simulation (\citealt{Pawlik2009, RaicevicTheuns2010}).
Larger values of the clumping factor should then only rule out late
reionisation with greater significance for models which have the same
$z_{\rm r}$.

\subsection{Prospects for constraints from future temperature measurements}
\label{SubSec:Upcoming}

In the previous section we presented constraints on the reionisation
redshift from the existing measurements of the IGM temperature around
quasars at $z\simeq 6$ (\citealt{Bolton2011a}). This work demonstrates
the potential of high redshift IGM temperature measurements to aid in
ruling out very late reionisation, so it is therefore interesting to
consider if future progress can be made with further data.

We begin by creating mock catalogues of possible future
observations. To do this, we assume that for a given $z_{\rm r}$,
ionising source model and density bias model, $N_{\rm obs}$
observations are to be drawn. For each one, the quasar spectral index
is drawn from the uniform prior over $\alpha_{\rm q}$ and the quasars
are assumed to have an optically bright phase which starts at a
redshift drawn from uniform distribution in the range $5.8 < z < 6.5$.
The median of each new observation is taken at random from our
predicted temperature distribution for the chosen $z_{\rm r}$, the
redshift at which the quasar turns on and $\alpha_{\rm q}$. Finally,
the errors in the measurement around this median are assumed to
correspond to one of the 14 (7 fiducial and 7 augmented Jeans scale)
observational measurements chosen at random. Once $N_{\rm obs}$
measurements are drawn in this way for each $z_{\rm r}$ and model, we
then use equations~(\ref{Eq:Bayes}) -- (\ref{Eq:BMA}) to obtain
predicted probability distributions.

This procedure for the generation of mock observations is repeated 100
times, and the resulting distributions are averaged. We then
marginalise over the different ionising source models to obtain a
predicted distribution $p(z_{\rm r})$ for every $z_{\rm r}$ from which
the measurements are drawn. This allows us to estimate the confidence
with which the reionisation redshift can be measured as a function of
the true $z_{\rm r}$. For simplicity, we omit the constraints imposed
by ${\tau_{\rm CMB}}$ in order to assess the new information drawn
from the temperature observations alone.

\begin{figure*}
\begin{center}
 \includegraphics[width=1.0\textwidth]{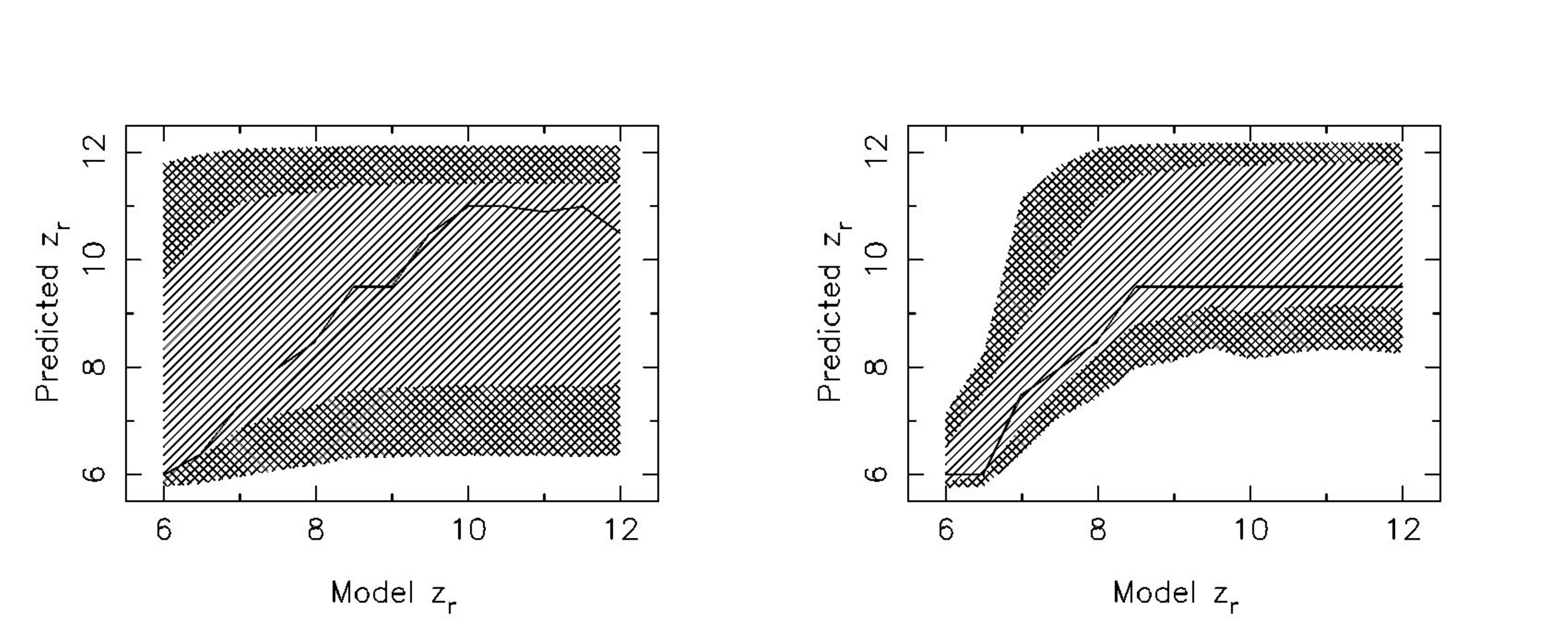}
\vspace{-0.3cm}
\caption{Bias and uncertainty in the constraints on the reionisation
  redshift from near-zone temperatures, as a function of the true
  reionisation redshift in our models. In each case the mode of the
  predicted posterior distribution is shown (black line) alongside the
  68 per cent (lined region) and 95 per cent (hatched region)
  confidence intervals.  {\em Left:} Results assuming 7 temperature
  measurements {\em Right:} Results for 100 near-zone temperature
  measurements, illustrating the optimistic case of large numbers of
  quasar sight-lines.}
\label{Fig:Confidences}
\end{center}
\end{figure*}

We present our results in Figure \ref{Fig:Confidences}. The left panel
shows the mode (solid line) and confidence intervals (filled regions)
for $N_{\rm obs} = 7$, mimicking the constraints of the previous
section.  Meanwhile, the right panel shows the same results for a
large number of observations\footnote{Note that a total of 100 quasar
  high-resolution spectra at $z \simeq 6$ is a very optimistic
  scenario.  Obtaining one $R\sim 40\,000$ spectrum, even for the
  brightest known quasars at these redshifts, requires $\sim 10$ hours
  of integration on an 8-m class telescope.}, $N_{\rm obs} = 100$.  We
note that, in both cases, the mode typically recovers the true
reionisation redshift in the models well, particularly for small
$z_{\rm r}$. However, at no reionisation redshift do we obtain a
particularly tight constraint for seven sight-lines only.  For late
reionisation, this is primarily due to uncertainties in modelling of
both the quasar spectrum and the ionising efficiency.  Similarly, if
reionisation happened early, we will not be able to distinguish well
between different $z_{\rm r}$ as the IGM temperature reaches an
asymptotic state independent of when reionisation occurred. However,
increasing the number of observations from 7 to 100 can tighten our
constraints on late reionisation considerably, increasing the
potential 95 per cent limit on $z_{\rm r}$ from 6.5 to 8.0.

Improved constraints the average quasar EUV spectra index from low
redshift data could help to tighten our prior on $\alpha_{\rm q}$, and
further improve limits at least on late reionisation.  However, to
tighten constraints on early reionisation using IGM temperature
measurements ultimately requires observations which push to even
higher redshifts.


\section{Conclusions}
\label{Sec:Conclusions}

Current understanding of the reionisation history is primarily built
on existing observational evidence comprised of measurements of the
Gunn-Peterson optical depth at $z \la6$, and the electron scattering
optical depth of cosmic microwave photons.  These measurements
respectively probe only the tail end of the epoch of reionisation, and
an integral measure of the reionisation history. The available
constraints therefore permit a wide range of possible
histories. Observations of the IGM temperature provide an additional
complementary probe since the thermal memory of the IGM yields
information at redshifts beyond the those where observations are made.
Measurements of the temperature around high redshift quasars therefore
offer a renewed opportunity to constrain the reionisation history of
the Universe.

In this paper, we have used measurements of IGM temperature at $z
\simeq 6$ (\citealt{Bolton2011a}) to constrain the redshift at which
the hydrogen reionisation epoch ended. Our analysis employs a
semi-numerical model for reionisation and heating, which is
constrained to fit the CMB electron scattering optical depth, and the
ionisation rate inferred from the Ly$\alpha$ forest at $z = 6$. We
find that all models which reionise the Universe prior to $z_{\rm r}
\approx 9.0$, settle to approximately the same IGM thermal state by $z
= 6.0$ with relatively little scatter in the temperature at mean
density.  For late reionisation, ($z_{\rm r} < 8.0$), the situation is
different since there is significantly more variance in the IGM
temperature distribution. This arises partly from differences in
ionisation time, which allow for different cooling times and a greater
scatter. However, we find that there is always a detectable thermal
imprint of at least $2\,500{\rm K}$, which differentiates the mean
temperature signifying late reionisation $z_{\rm r} \approx 6$ from the asymptotic state
achieved for $z_{\rm r} > 9.0$. This implies that measurements of
temperature in quasar near-zones can be used to constrain the
reionisation redshift if it is sufficiently late.

Including observational constraints from the CMB electron scattering
optical depth and the \Lya forest opacity, we conclude that $z_{\rm r}
> 7.9$ $(6.5)$ with 68 (95) per cent confidence.  The $z\simeq 6$
temperature measurements therefore rule out a very late completion to
reionisation.  The inclusion of further temperature measurements in
the future will tighten these constraints.  If reionisation occurred
early, we estimate a sample of 100 quasar spectra would rule out
$z_{\rm r} < 8.0$ to 95 per cent confidence.  However, in order to
push these constraints further, more detailed modelling of the IGM
thermal evolution will also be required.  In particular, the evolution
of the ionising efficiency with redshift will have to be understood in
order to narrow the modelling uncertainties for later reionisation
scenarios.  Including the effect of hydrogen reionisation sources
which have harder spectra than we have assumed here will also be
valuable.  Finally, improved estimates of quasar spectral indices will
aid in reducing uncertainties in the modelling of the near-zone
temperatures.

\section*{Acknowledgments}

SR would like to thank the Institute of Astronomy in Cambridge and in
particular Martin Haehnelt for their hospitality and support.  JSB and
JSBW acknowledge the support of the Australian Research Council. The
Centre for All-sky Astrophysics is an Australian Research Council
Centre of Excellence funded by grant CE11E0090.  GDB thanks the Kavli
foundation for financial support.

\end{document}